\documentclass[draftclsnofoot, 10 pt, onecolumn]{IEEEtran}

\IEEEoverridecommandlockouts
\let\proof\relax
\let\endproof\relax
\usepackage{amsmath, amsthm, amssymb, amsfonts, mathtools, xargs, tensor, units, cite, stmaryrd, mathrsfs, algpseudocode, comment, graphicx}
\usepackage{float}
\usepackage{algorithm}
\usepackage{multibib}
\usepackage[unicode=true, bookmarks=true, bookmarksnumbered=false, bookmarksopen=false, breaklinks=false, pdfborder={0 0 1}, backref=false, colorlinks=false, hidelinks]{hyperref}
\makeatletter
\let\NAT@parse\undefined
\makeatother

\usepackage[usenames, dvipsnames]{color}

\usepackage[bytype]{Theorems}

\usepackage{geometry}
\title{Carleman Lifting for Nonlinear System Identification with Guaranteed Error Bounds}
	
	\author{Moad Abudia\textsuperscript{1}, Joel A. Rosenfeld\textsuperscript{2}, and Rushikesh Kamalapurkar\textsuperscript{1} \thanks{\textsuperscript{1} The authors are with the School of Mechanical and Aerospace Engineering, Oklahoma State University, Stillwater, OK, USA. {\tt\small \{abudia, rushikesh.kamalapurkar\}@okstate.edu}.\\
	\textsuperscript{2} The author is with the Department of Mathematics and Statistics, University of South Florida, Tampa, Fl, USA. {\tt\small rosenfeldj@usf.edu}.This research was supported, in part, by the National Science Foundation (NSF) under award number 2027999, and the Air Force Office of Scientific Research under award number FA9550-20-1-0127. Any opinions, findings, conclusions, or recommendations detailed in this article are those of the author(s), and do not necessarily reflect the views of the sponsoring agencies.}}
\begin{document}


	\maketitle
\thispagestyle{empty}
\pagestyle{empty}
	\begin{abstract}
		This paper concerns identification of uncontrolled or closed loop nonlinear systems using a set of trajectories that are generated by the system in a domain of attraction. The objective is to ensure that the trajectories of the identified systems are close to the trajectories of the real system, as quantified by an error bound that is  \textit{prescribed a priori}. A majority of existing methods for nonlinear system identification rely on techniques such as neural networks, autoregressive moving averages, and spectral decomposition that do not provide systematic approaches to meet pre-defined error bounds. The developed method is based on Carleman linearization-based lifting of the nonlinear system to an infinite dimensional linear system. The linear system is then truncated to a suitable order, computed based on the prescribed error bound, and parameters of the truncated linear system are estimated from data. The effectiveness of the technique is demonstrated by identifying an approximation of the Van der Pol oscillator from data within a prescribed error bound.
	\end{abstract}
	

	\section{Introduction} \label{section:Introduction}
	
	Identifying nonlinear dynamical systems from data, without knowledge of the structure of the model, and with guaranteed error bounds, has proven to be a difficult challenge. The focus of this paper is on the identification of the dynamic model of a nonlinear system within prescribed error bounds using Carleman lifting, where the finite dimensional nonlinear system is lifted into an infinite-dimensional linear system via Carleman linearization. The linear system is then truncated to a suitable order, computed based on the prescribed error bound, and the parameters of the truncated linear system are estimated from data.

A variety of system identification methods are available for linear time invariant (LTI) systems. For example, in non-parametric frequency-domain estimation methods, the system is subjected to a white noise input, the output is represented in the frequency domain, and the user fits a model that is determined by the number of resonant frequencies and the decay rate of the power spectrum. On the other hand, parametric methods presuppose a system representation, such as an auto regressive moving average (ARMA) model or a state space (SS) model, where the order of the system is estimated first using auto correlation. The system identification problem is reduced to a parameter estimation problem, which is solved using numerical gradient descent methods to minimize a suitable error metric (c.f.\cite{SCC.Ljung1999}).

In the case of nonlinear systems, the identification problem is much more difficult. If the structure of the system is known \textit{a priori}, the identification problem can be reduced to a parameter estimation problem which can be solved using Lyapunov-based adaptive estimation methods \cite{SCC.Mahyuddin.ea2013}. In the case where the structure of the system is unknown, some assumptions have to be made. For instance, in \cite{SCC.Brunton.Proctor.ea2016a} a set of basis functions are specified beforehand, such as constant, polynomial, and trigonometric functions, in the hope that the data can be decomposed appropriately with respect to the chosen basis. Another approach is using kernel dynamic mode decomposition (DMD) (c.f. \cite{SCC.Williams.Rowley.ea2015}) which aims to decompose a time series corresponding to a nonlinear dynamical system into a collection of dynamic modes using spectral decomposition of the Koopman operator. In order to establish convergence guarantees and error bounds, the Koopman operator needs to be compact \cite{SCC.Mezic2013}, \cite{SCC.Pedersen2012}, which is typically not the case when the dynamics are nonlinear \cite{SCC.Gonzalez.Abudia.easubmitted}, although \cite{SCC.Korda.Mezic2018} shows convergence in the strong operator topology, convergence of the spectra is needed. In \cite{SCC.klus2020} the Koopman generator is aproximated from data, which can be used to compute the eigenvalues, eigenfunctions, and modes of the generator and for system identification without any error bound garuntees. While the techniques described above have been proven effective in many practical applications \cite{SCC.Brunton.Proctor.ea2016a},\cite{SCC.Williams.Rowley.ea2015}, \cite{SCC.Rosenfeld.Kamalapurkar.ea2019a}, they generally do not provide theoretical error bound guarantees. 

The approach in this paper is inspired by the Carleman Linearization approach developed in \cite{SCC.Carleman.1932}. Carleman linearization converts a finite dimensional nonlinear system to an infinite dimensional linear system. The conversion is realized by expanding the state space of the nonlinear system to include all monomial functions of the original state variables.

In \cite{SCC.Amini.ea2021}, the authors used Carleman Linearization in conjunction with a known nonlinear model to develop a truncated linear system that approximates the nonlinear system in a region of attraction. Moreover, the trajectory of the truncated linear system is guaranteed to stay within a computable error bound around the trajectory of the nonlinear system. The error bound can be computed \textit{a priori} when the decay rate of the Maclaurin expansion of the nonlinear system and the order of the truncation are known.

In this paper, the results in \cite{SCC.Amini.ea2021} are leveraged to develop a Carleman lifting-based approach to data-driven modeling of nonlinear systems. Given a set of trajectories generated by a nonlinear system and an upper bound of the decay rate of the Maclaurin expansion, a linear system of a higher order is identified such that the trajectory of the identified linear system stays within a guaranteed error bound around the trajectory of the nonlinear system, and the error bound can be \textit{prescribed a priori}.

The paper is organized as follows. The problem is formulated in Section \ref{section:Problem Statement}. In Section \ref{section:Data-driven_Carleman_Lifting} the approach for identification of linear systems is presented. In Section \ref{section:Error Analysis and Prescribed Error Approximation} the properties of the identified system are analyzed. In Section \ref{Order Selection Algorithm} an algorithm to generate the identified linear system with a guaranteed error bound is presented. In Section \ref{section:Simulation} a simulation is presented to demonstrate the system identification method. In Section \ref{section:Discussion}, a discussion of the results is presented, and Section \ref{section:Conclusion and future work} includes concluding remarks and a discussion on future work.

	\section{Problem Statement} \label{section:Problem Statement}
	Consider an unknown dynamical system of the form
\begin{equation}
\label{orginal_system}
    \dot{x}=\mathbf{f}(t,x)
\end{equation}
where $\mathbf{f}:\left(\mathbb{R}_{+} \times \mathbb{R}^{d}\right) \rightarrow \mathbb{R}^{d}$ is a vector valued real analytic function of several variables, $\mathbb{R}_{+} \coloneqq\{y\in\mathbb{R}:y\geq 0\}$,    $x(0)=x_0$, and $f(t,0)=0$. Given a set of observed trajectories, $\{\gamma_i\}_{i=1}^m=X$, generated from \eqref{orginal_system} and an error bound $\Delta > 0$, the objective is to develop a systematic technique to either 
\begin{enumerate}
    \item construct another dynamical system of the form $\dot{\hat{z}} = g(t,\hat{z})$ such that the error $\left \|x(t)-\hat{z}\mid_d(t)\right \|$ between the trajectories of the $\hat{z}-$system and the trajectories of \eqref{orginal_system} is less than $\Delta$ for $t\in[0,\tau^{*}]$, for some $\tau^* > 0$, where $\hat{z}\mid_d$ is the truncation of $\hat{z}$ to the first $d$ dimensions, or
    \item conclude that construction of such a system, using the particular method developed in this paper, is not possible.
\end{enumerate}

The problem, as formulated above, is difficult to solve for general nonlinear systems. In this paper the formulation is restricted to a sub-class of nonlinear systems, defined by the following assumptions as in \cite{SCC.Amini.ea2021}.

\begin{assumption}\label{as:1}
The vector field $\mathbf{f}(t,x)$ admits a Maclaurin expansion about the state vector $x$ 

\begin{equation}
\mathbf{f}(t, x)=\sum_{\boldsymbol{\alpha} \in \mathbb{Z}_{+}^{d}} \mathbf{f}_{\boldsymbol{\alpha}}(t) \mathbf{x}^{\boldsymbol{\alpha}}=\sum_{\boldsymbol{\alpha} \in \mathbb{Z}_{+}^{d} \backslash\{0\}} \mathbf{f}_{\boldsymbol{\alpha}}(t) \mathbf{x}^{\boldsymbol{\alpha}}, t \in \mathbb{R}^{+}.
\end{equation}
where $\boldsymbol{\alpha}=\left(\alpha_1, \ldots, \alpha_d\right) \in \mathbb{Z}_{+}^d$, is a multi-index and its
corresponding multi-variate monomial is $\mathbf{x}^\alpha=x_1^{\alpha_1} \cdots x_d^{\alpha_d}$, where $\mathbb{Z}_{+}^d$ is the set of non-negative integers.
\end{assumption}

\begin{assumption}\label{as:2}
The Maclaurin expansion coefficients satisfy the exponential decay property

\begin{equation}
\label{decay_rate}
\sup _{t \geq 0} \sum_{|\alpha|=n}\left\|\mathbf{f}_{\boldsymbol{\alpha}}(t)\right\|_{\infty} \leq C R^{-n}, n \geq 0
\end{equation}
where $C$ and $R$ are positive constants, and the cardinality of $\boldsymbol{\alpha}$ is defined by $|\boldsymbol{\alpha}|=\alpha_1+\cdots+\alpha_d$.
\end{assumption}
While this assumption is restrictive, a large subclass of systems fall under this category, since polynomial, trigonometric, and exponential functions have Maclaurin expansions with exponential decay.
The idea is to use Carleman lifting \cite{SCC.Carleman.1932},\cite{SCC.Amini.ea2021} to lift the nonlinear system into an infinite-dimensional linear system such that the truncation of the trajectory of the linear system to the first $d$ dimensions approximates the trajectory of the system in (\ref{orginal_system}), in the infinity norm over a finite time interval. The infinite-dimensional linear system is then truncated to yield a finite-dimensional linear system such that the error between the projected trajectories and the original trajectories is less than the given error bound, $\Delta$.
	
	\section{Data-driven Carleman Lifting} \label{section:Data-driven_Carleman_Lifting}
 Carleman linearization \cite{SCC.Carleman.1932} lifts a finite dimensional nonlinear system to an infinite dimensional linear system
 \begin{equation}
     \dot{y}= \mathcal{A} y
 \end{equation}
 where $y$ is an infinite dimensional vector consisting of all unrepeated monomials of $x$ and $\mathcal{A}$ is an infinite dimensional operator.
 The operator $\mathcal{A}$ is approximated by truncating its matrix representation to the first $\mathbb{M}\times\mathbb{M}$ block, which will be called $A$.

	The nonlinear system is then approximated by a linear system in terms of a lifted state $z$ which consists of unrepeated monomials of $x$ up to order $N$. For example, if $N=3$ then
\begin{equation*}
    z=l(x) = \begin{array}{c}
        [x_{1}, \ldots,x_{d}, x_{1}^{2}, x_{1} x_{2},\ldots, x_{1} x_{d}, x_{2}^2, \ldots, x_{d}^{2},\\ 
         x_{1}^{3}, x_{1}^{2} x_{2}, \ldots, x_{1} x_{2} x_{3}, \ldots, x_{d}^{3} ]^{T},
       
    \end{array}
\end{equation*}
where $l:\mathbb{R}^d\to\mathbb{R^M}$ denotes the lifting map and $\mathbb{M}$ is the number of monomials which is $\mathbb{M}=\sum_{k=1}^{N} \genfrac(){0pt}{2}{k+d-1} {d-1}$.

The results of \cite{SCC.Amini.ea2021} indicate that for any system that satisfies Assumptions \ref{as:1} and \ref{as:2}, a trajectory in a region of attraction can be approximated, with arbitrary accuracy, by truncation of solutions of 
\begin{equation}
\label{lifted_system}
    \dot{z}=Az 
\end{equation}
to the first $d$ dimensions, where computation of $A$ requires complete knowledge of the system dynamics, $\mathbf{f}$.

In this paper, a data-driven approach to generate the linear system is developed. Given a set of trajectories of the nonlinear system, denoted by $\{\gamma_i\}:[0,T]\to\mathbb{R}^d$, that satisfy $\left\Vert \gamma_i(t) \right\Vert \leq M$ for all $t\in[0,T]$, the objective is to find an order $N$ and an estimate $\hat{A}$ of an $N$\textsuperscript{th}-order truncation $A$ of the operator $\mathcal{A}$ such that $ \sup_{t\in[0,\tau^*]}\left\|x(t)-\hat{z}\mid_d(t)\right \| \leq \Delta$, where $[0,\tau^*]$ is the interval over which the error bound can be guaranteed and $\hat{z}\mid_d$ is the $d-$dimensional truncation of the solution $\hat{z}$ of $\dot{\hat{z}} = \hat{A}\hat{z}$ starting from $\hat{z}(0) = l(x(0)) $.

Let $z_i$ denote the trajectory of (\ref{lifted_system}) starting from the initial condition $z_i(0) = l(\gamma_i(0))$. Let 
\begin{equation}
    \label{carleman_error_full}
\epsilon_i(t) \coloneqq l(\gamma_i(t)) - z_i(t)  
\end{equation}
denote the error between the trajectories of the model-based Carleman linearization in (\ref{lifted_system}) and the trajectories of (\ref{orginal_system}). The trajectories of the linear and the nonlinear systems are then related by
\begin{equation}
\label{lifted_system_with_error}
    \frac{d}{dt}l(\gamma_i(t)) = \dot{z_i}(t) + \dot{\epsilon}_i(t) = Az_i(t) + \dot{\epsilon}_i(t).
\end{equation}
Integrating (\ref{lifted_system_with_error}),
\begin{align}
    \int_0^T \frac{d}{dt}l(\gamma_i(t)) &= \int_0^T Az_i(\tau)d\tau + \epsilon_i(T)\nonumber\\ &= A\int_0^T \left(l(\gamma_i(\tau)) - \epsilon_i(\tau)d\tau\right) + \epsilon_i(T).
\end{align}
Note that $\epsilon_i(0) = 0$. Concatenating all the measured trajectories into a vector $\Gamma(t) = [l(\gamma_1(t))\,,\hdots,l(\gamma_m(t))]^T$, and letting $ I = \int_0^T Z(\tau)d\tau $, where $Z=\left[z_{1}, \ldots, z_{m}\right]$, with $ I_\epsilon = \int_0^T \epsilon(\tau)d\tau $ and $ I_{\Gamma} = \int_0^T \Gamma(\tau)d\tau $, one gets the relationships $ I_{\Gamma} = I + I_{\epsilon} $ and
\begin{equation}
\label{lifted_system_integrated}
    \Gamma(T) - \Gamma(0) = AI_\Gamma - AI_\epsilon + \epsilon(T),
\end{equation}
where $\epsilon(t) = [\epsilon_1(t)\,,\hdots,\epsilon_m(t)]^T$.

Provided the matrix $I_\Gamma$ is full rank, a data-driven approximation $\hat{A}$ of the lifted matrix $A$ can be computed using the least-squares solution of (\ref{lifted_system_integrated}) as
\begin{equation}
\label{A-determination}
   \hat{A} = \left(\Gamma(T)-\Gamma(0)\right) I_\Gamma^\dagger,
\end{equation}
where $(\cdot)^\dagger$ denotes the Moore-Penrose pseudo-inverse.

Once the matrix $\hat{A}$ is computed from data, trajectories of the linear system $\dot{\hat{z}} = \hat{A}\hat{z}$ can be computed and truncated to the first $d$ dimensions to estimate the trajectories of the nonlinear system. 
	
	\section{Error Analysis and Prescribed Error Approximation} \label{section:Error Analysis and Prescribed Error Approximation}
	
	There are two sources of error between the trajectories of the identified system truncated to the first $d$ dimensions, and the trajectories of the nonlinear system. The first, is the error $\epsilon$ introduced in (\ref{carleman_error_full}), between the trajectories of the nonlinear system and the model-based Carleman linearized system in (\ref{lifted_system}). The second, is the error $\tilde{z}(t) \coloneqq z(t) - \hat{z}(t) $, between the trajectories of the model-based Carleman linearized system and the identified linear system, starting from the same initial conditions. The former is quantified in \cite{SCC.Amini.ea2021} as follows. 
	\begin{assumption}\label{as:3}
	The measured trajectories are bounded such that $\|\gamma_i(t)\|\leq M$ for all $i=1,...,d$ and $t\in [0,T]$ and $$M<\frac{R}{e},$$
	where $R$ is introduced in (\ref{decay_rate}) and $e$ is the base of the natural logarithm.
	\end{assumption}
	Under Assumptions \ref{as:1} - \ref{as:3}, for every $N\geq 1$ there exists a $\tau^{*}>0$ such that
\begin{equation}
    \label{carleman_error_bound}
    \left \|\epsilon\mid_d(t)\right \| \leq D\mu^{N},\forall t\in [0,\tau^{*}],
\end{equation}
where $\epsilon\mid_d$ is the $d-$dimensional truncation of $\epsilon$ in (\ref{carleman_error_full}), $$
0<D=M\left(1-\left(\frac{M}{R}\right)\right)^{-1}
,$$ 
  $$C_{0}\leq CR^{-1},$$ and
$$
\mu<\left(M e R^{-1}\right) e^{C_{0} \tau^{*}}<1.
$$
The error bound in (\ref{carleman_error_bound}) can then be guaranteed over the time interval $[0,\tau^{*}]$  where
\begin{equation} \label{tau_bound}
    \tau^{*}<\frac{-\log(MeR^{-1})}{C_0},
\end{equation}

One of the main contribution of this paper is to quantify the error $\tilde{z}$, between the trajectories of the identified system and the model-based Carleman linearized system. 
\begin{theorem}
If assumptions \ref{as:1},\ref{as:2},\ref{as:3}, and \ref{as:4} hold, then the state estimation error $\left\|x(t)-\hat{z}\mid_d(t)\right \| \leq D\mu^N+t\overline{B} \overline{z} \overline{A}, \hspace{1em}\forall t\in [0,\tau^{*}]$ for $\overline{B}, \overline{z}$, and $\overline{A}$  from (\ref{z_tilde bound}) as described in the following.
\end{theorem}

\begin{proof}

The dynamics of the error are given by  
	   $$
	   \dot{\tilde{z}}=A z-\hat{A}\hat{z}.
	   $$
By adding and subtracting $A\hat{z}$,
	   $$
	   \dot{\tilde{z}}=A \tilde{z}+(A-\hat{A})\hat{z}.
	   $$
Using the variation of constants formula \cite{SCC.coddington2012}
	   $$
	   \tilde{z}(\tau^*)=e^{A\tau^*}\tilde{z}(0)+\int_0^{\tau^*} e^{A(\tau^*-\tau)}(A-\hat{A})\hat{z}(\tau) d\tau.
	   $$
	   The triangle inequality and the
Cauchy-Schwarz inequality then result in the bound
	   $$
	    \left\|\tilde{z}(\tau^*)\right\| \leq \int_0^{\tau^*}\left\|e^{A(\tau^*-\tau)}\right\| \left\|A-\hat{A}\right\|
	    \left\|\hat{z}(\tau)\right\| d\tau .
	   $$
 Since $\tilde{z}(0)=0$, then 
\begin{equation}
   	   \left\|\tilde{z}(\tau^*)\right\| \leq \tau^*\overline{B} \overline{z} \overline{A}
\label{z_tilde bound}
\end{equation}

 where $\overline{B}=\max{\{\left\|e^{A\tau^*}\right\|,1\}}$,  $\overline{z}= \sup_{t\in [0,\tau^*]} \left\|\hat{z}(t)\right\|$, and $ \overline{A}=\left\|A-\hat{A}\right\| $.

 Since $A$ is unknown, calculation of $\overline{A}$ requires further effort. To that end, let
	   $$
	   D(t)=Z(t)-Z(0).
	   $$
 From (\ref{lifted_system}) if $II^T$ is nonsingular, then 
	   $$
	   A=DI^T(II^T)^{-1}, 
	   $$
	   and from (\ref{A-determination}) the estimation is calculated as   
	   $$
	   \hat{A}=(D+\epsilon) (I+I_{\epsilon})^T[(I+I_{\epsilon})(I+I_{\epsilon})^T]^{-1}.
	   $$
To compute a bound on $\tilde{z}$, the following assumption is needed.
\begin{assumption}\label{as:4}
	There exists a constant $\overline{I}$ such that $\left\|(I^TI)^{-1}\right\|\leq \overline{I}$.
\end{assumption}
Note that the matrix $I$ is comprised of integrals of the trajectories of the model-based Carleman linearized system, and as such, it is unknown. However, the matrix $I_\Gamma$ can be computed, and is perturbed from $I$ by $I_\epsilon$. Using continuity of eigenvalues of matrices with respect to elements of the matrix, it can be concluded that for small enough $I_\epsilon$, the difference between $\left\|(I^T I)^{-1}\right\|$ and $\left\|(I_\Gamma^T I_\Gamma)^{-1}\right\|$ is $o(I_\epsilon)$.

The bound $\overline{A}$ can then be estimated using the matrix inverse identity \cite{SCC.Henderson.Searle}, 
$$
(Q+V)^{-1}=Q^{-1}-Q^{-1}V(Q+V)^{-1},
$$
where $Q$ and $(Q+V)$ are non-singular matrices. Applying the identity to the expression for $\hat A$,  
\begin{multline}
\label{A_bar}
	  \overline{A} \leq
	    \left\|\overline{I}\right\| (\left\|I^T\right\| \left\|\epsilon\right\|+\left\|I^T_\epsilon\right\|\left\|D\right\|+\left\|I^T_\epsilon\right\| \left\|\epsilon\right\|) \\+
	     [\left\|\overline{I}\right\|(\left\|I^T\right\| \left\|I_\epsilon\right\|+\left\|I_{\epsilon}^{T}\right\|\left\|I\right\|+\left\|I_\epsilon\right\|\left\|I^T_\epsilon\right\|)]\\\cdot [\left\|I^T\right\|\left\|D\right\|+\left\|I^T\right\|\left\|\epsilon\right\|+\left\|I^T_\epsilon\right\|\left\|D\right\|+\left\|I^T_\epsilon\right\|\left\|\epsilon\right\|],
\end{multline}
where $\left\|I^T\right\| \leq \left\|I_\Gamma^T\right\|+\left\|I_\epsilon^T\right\|$.\\

Since $\overline{B}=\max\{\left\|e^{A\tau^*}\right\|,1\}$, an estimate of $\left\|e^{A\tau^*}\right\|$ is obtained by realizing that
$$
e^{A\tau^*}=e^{(A-\hat{A}+\hat{A})\tau^*}=e^{A-\hat{A}\tau^*}e^{\hat{A}\tau^*},
$$
and as a result,
\begin{equation}
    \label{B_bar}
\overline{B}=\left\|e^{A\tau^*}\right\| \leq e^{\overline{A} \tau^*}\left\|e^{\hat{A}\tau^*}\right\|.
\end{equation}\\
By using the triangle inequality
	  $$
	  \left\|x(t)-\hat{z}\mid_d(t)\right \| \leq \left\|\epsilon \mid_d(t)\right \|+\left\|\tilde{z}\right \|,
	  $$
using (\ref{z_tilde bound}) and (\ref{carleman_error_bound})  it can be further simplified as 
	  \begin{equation}
\label{error_bound}
	  \left\|x(t)-\hat{z}\mid_d(t)\right \| \leq D\mu^N+t\overline{B} \overline{z} \overline{A}, \hspace{1em}\forall t\in [0,\tau^{*}].
	  \end{equation} 
\end{proof}

   \section{Order Selection Algorithm} \label{Order Selection Algorithm}
	  Once the upper bound is calculated as a function of $N$ using (\ref{error_bound}), a search over $N=1,2,\hdots,\overline{N}$ can be conducted (see Algorithm \ref{alg:cap}), to yield
	  $$
	  N^*= \arg \min_N(D\mu^N+\tau^*\overline{B} \overline{z} \overline{A}),
	  $$
	  where $N^*$ denotes the lifted order which produces an identified system with the smallest guaranteed error bound. If $D\mu^{N^*}+\tau^*\overline{B} \overline{z} \overline{A} > \Delta$, then the system identification method cannot produce a system that guarantees $\sup _{t \in [0,\tau^*]}\left \|x(t)-\hat z\mid_2(t)\right \| <\Delta$. Otherwise, if $D\mu^{N^*}+\tau^*\overline{B} \overline{z} \overline{A} \leq \Delta$, then a system can be identified using the truncation order $N^*$ to guarantee $\sup _{t \in [0,\tau^*]}\left \|x(t)-\hat z\mid_2(t)\right \| <\Delta$.

\begin{algorithm}
\caption{Carleman System identification algorithm. In the algorithm, $\overline{N}$ is an upper bound on the truncation order, selected a priori.}\label{alg:cap}
\begin{algorithmic}
\Require $X$, $R$, $C$, $\overline{N}$ and $\Delta$
\Ensure $C_{0}\leq CR^{-1}$
\Ensure $M>||x||_\infty$

\State $D \gets M\left(1-\left(\frac{M}{R}\right)\right)^{-1}$
\Ensure $D>0$
\Ensure $\tau^{*}<\frac{-\log(MeR^{-1})}{C_0}$
\Ensure $\mu<\left(M e R^{-1}\right) e^{C_{0} \tau^{*}}<1$
\State $N \gets 1$
\While{$N \neq \overline{N}$}
\State Find $\overline{A}$ from (\ref{A_bar}) 
\State Find $\overline{B}$ from (\ref{B_bar}) 
\State $\Theta(N) \gets D\mu^N+\tau^*\overline{A}\overline{z}\overline{B}$
\State $N \gets N+1$
\EndWhile
\If{$\min[\Theta(N)] \leq \Delta$}
    \State $N^* \gets \arg \min_N [\Theta(N)]$
    \State Return $\hat{A}$ using $N^*$
\ElsIf{$\min[\Theta(N)] > \Delta$}
    \State Return ``Failed"
\EndIf

\end{algorithmic}
\end{algorithm}

\begin{figure*}[htp]
    \centering
        \begin{minipage}[t]{0.49\textwidth}
        \includegraphics[width=\textwidth]{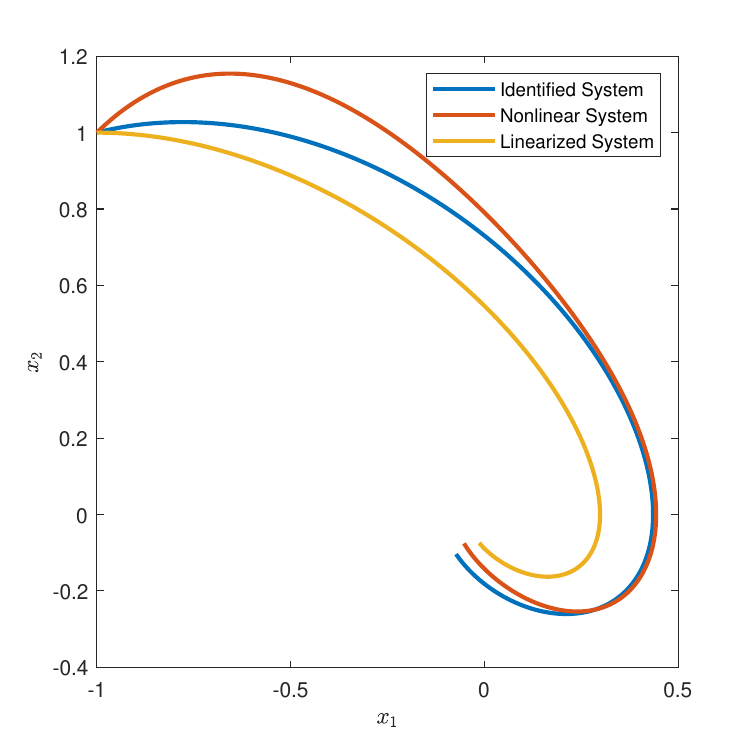}
        \caption{Trajectories of the nonlinear system, the identified system, and the model-based Carleman linearized system using $N=2$. } \label{fig:order_2}
    \end{minipage}\hfill
    \begin{minipage}[t]{0.49\textwidth}
        \includegraphics[width=\textwidth]{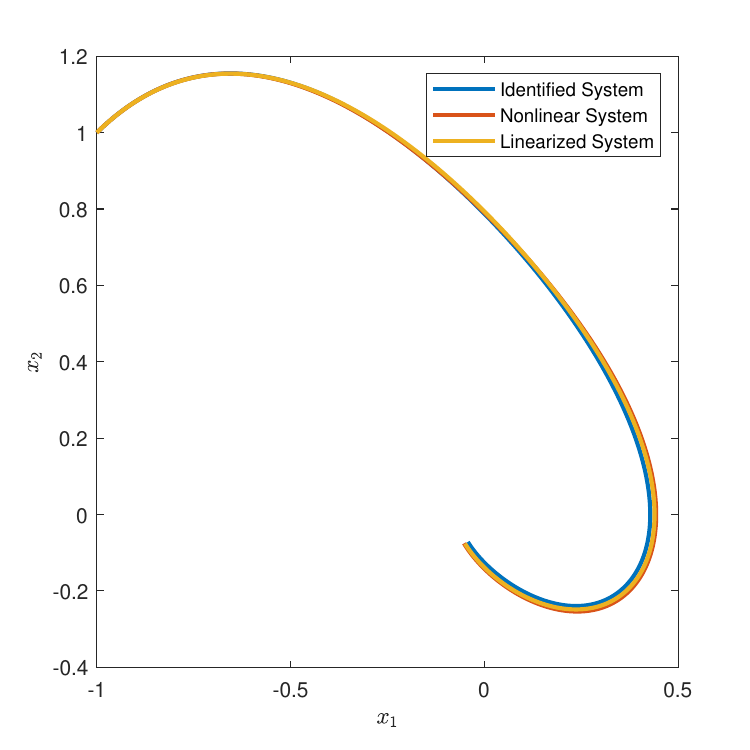}
        \caption{Trajectories of the nonlinear system, the identified system, and the model-based Carleman linearized system using  $N=5$. } \label{fig:order_5}
    \end{minipage}
\end{figure*}

\begin{figure}[htp]
    \centerline{\includegraphics[width=0.49\textwidth]{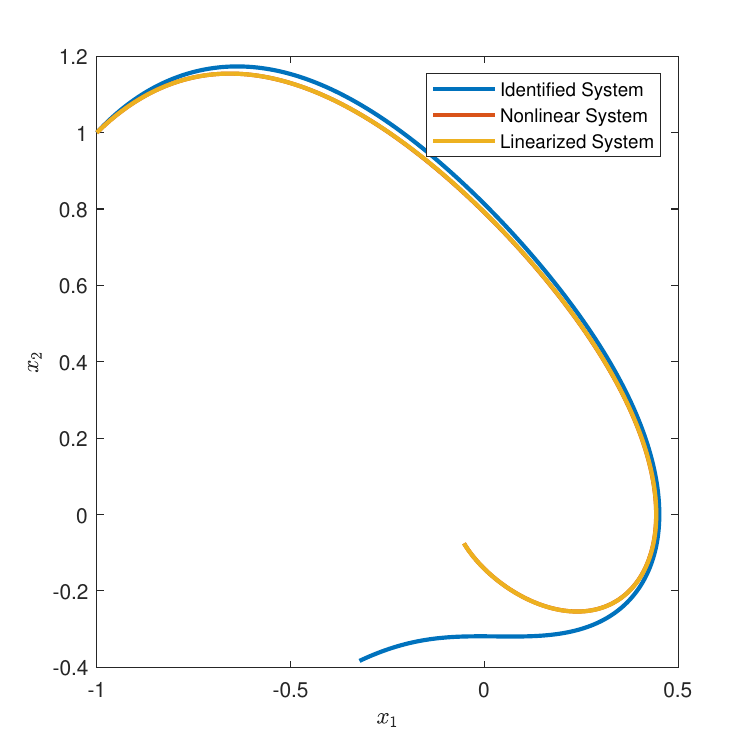}}
        \caption{Trajectories of the nonlinear system, the identified system, and the model-based Carleman linearized system using $N=11$. } \label{fig:order_11}
\end{figure}

\begin{figure*}[htp]
    \centering
        \begin{minipage}[t]{0.49\textwidth}
        \includegraphics[width=\textwidth]{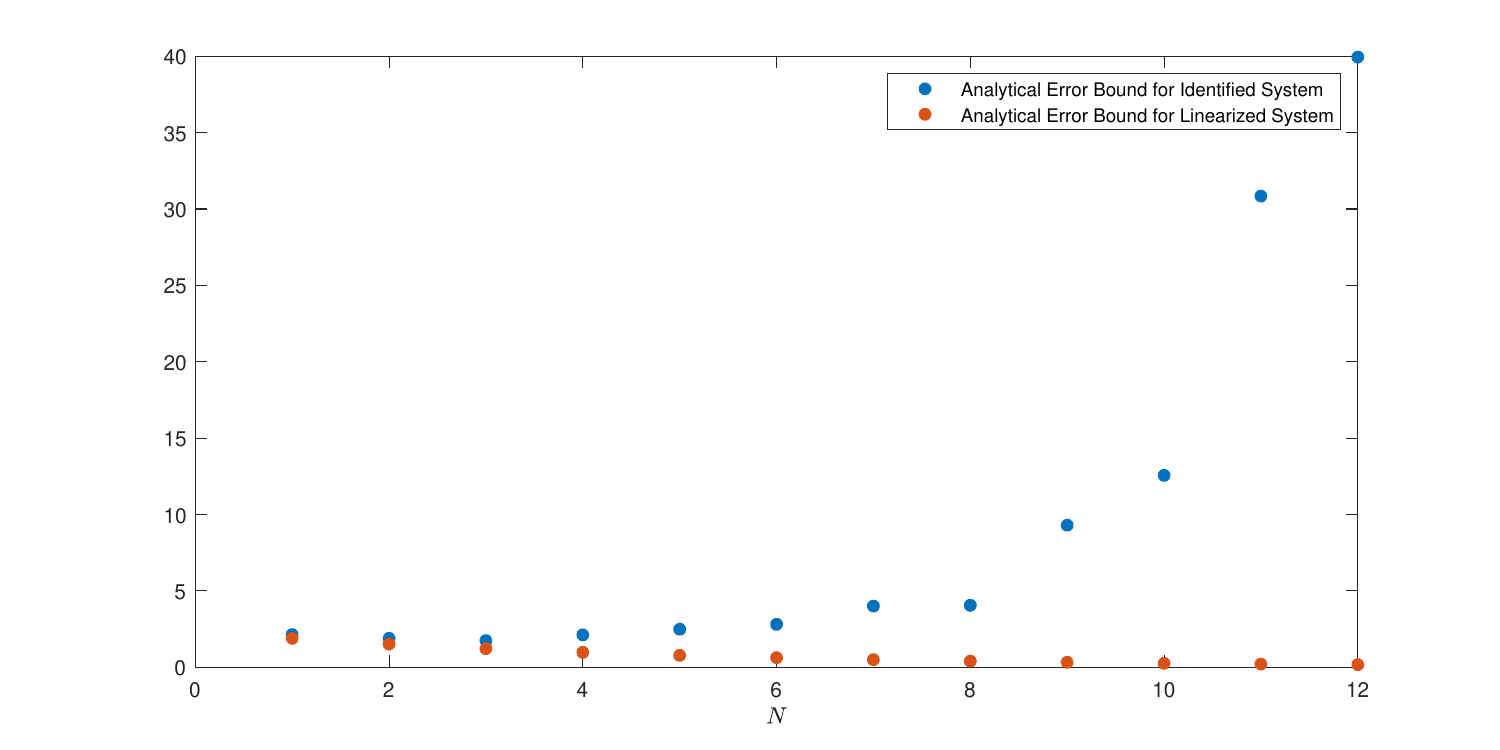}
        \caption{Comparison of analytical errors bounds on $\sup _{t \in [0,\tau^*]}\left \|x(t)-\hat z\mid_2(t)\right \|$, the error between the trajectories of the nonlinear system and the identified system, and $\sup _{t \in [0,\tau^*]}\left \|x(t)- z\mid_2(t)\right \|$, the error between the trajectories of the nonlinear system and the model-based Carleman linearized system, for truncation orders $N=1,2,...,12$. } \label{fig:error_bound}
    \end{minipage}\hfill
    \begin{minipage}[t]{0.49\textwidth}
        \includegraphics[width=\textwidth]{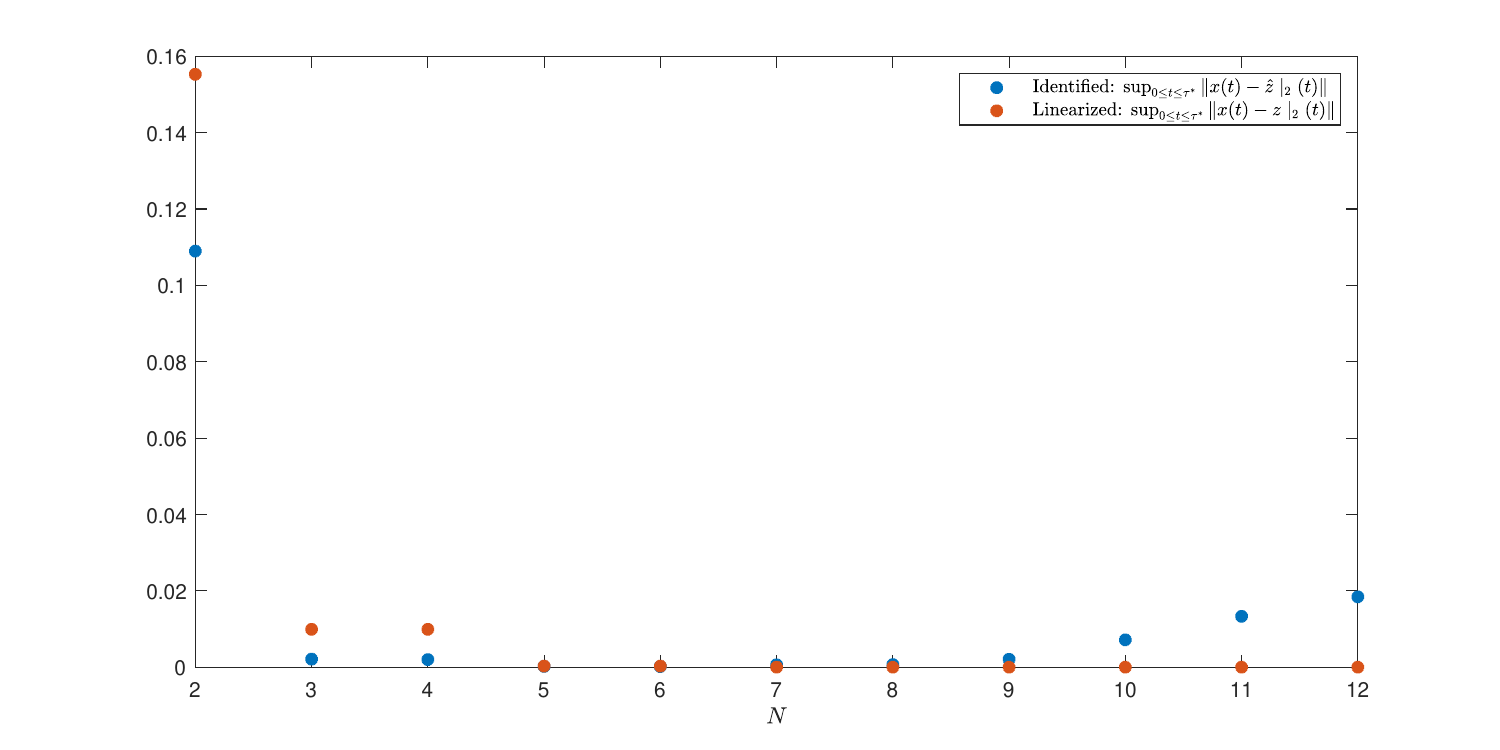}
        \caption{Comparison of errors $\sup _{t \in [0,\tau^*]}\left \|x(t)-\hat z\mid_2(t)\right \|$ between the measured trajectories and the identified trajectories, and $\sup _{t \in [0,\tau^*]}\left \|x(t)- z\mid_2(t)\right \|$, between the measured trajectories and the model-based Carleman linearized trajectories, for truncation orders $N=1,2,...,12$. } \label{fig:error_plot}
    \end{minipage}
\end{figure*}

\begin{figure*}[!ht]
    \centering
        \begin{minipage}[t]{0.49\textwidth}
        \includegraphics[width=\textwidth]{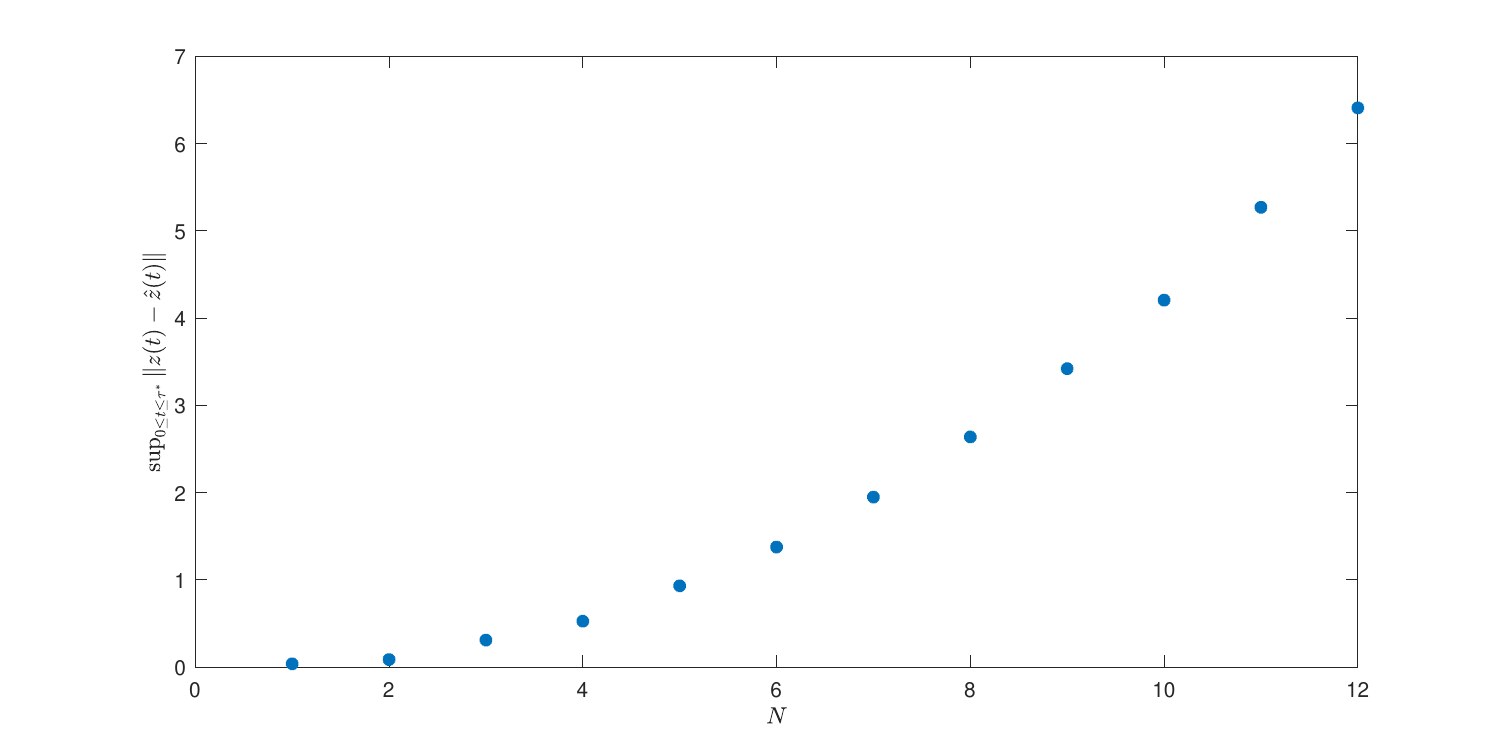}
        \caption{maximum error norm of the full state between the linearized system and the identified system  using $N=1,2,...,12$  } \label{fig:error_plot_full_z_zhat}
    \end{minipage}\hfill
    \begin{minipage}[t]{0.49\textwidth}
        \includegraphics[width=\textwidth]{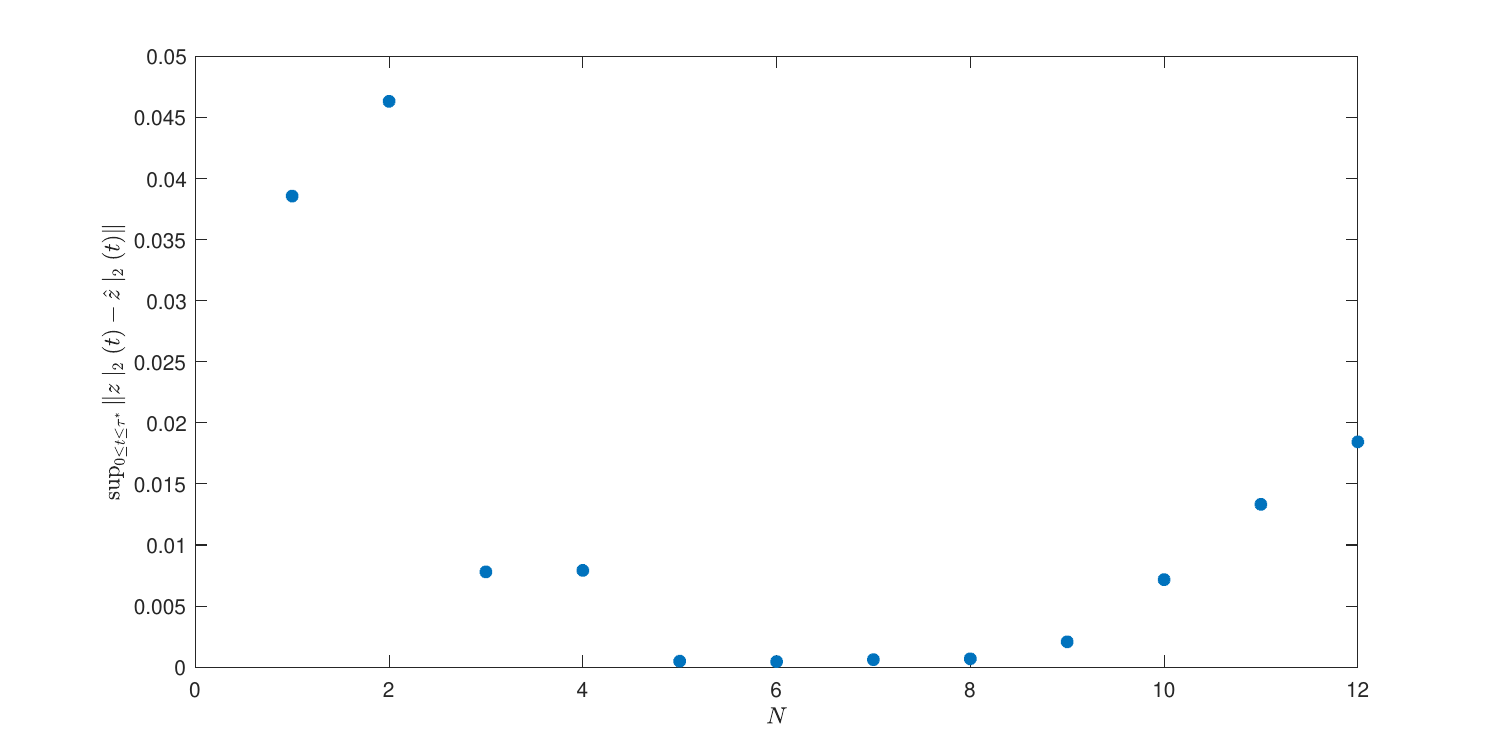}
        \caption{maximum error norm of the first two states between the linearized system and the identified system  using $N=1,2,...,12$ } \label{fig:error_plot_1_2_z_zhat}
    \end{minipage}
\end{figure*}

	\section{Simulation}
	\label{section:Simulation}
To demonstrate the effectiveness  of the system identification approach in Section \ref{section:Data-driven_Carleman_Lifting}, the Van der Pol Oscillator is used. Consider the Van der Pol Oscillator given by
\begin{equation}
\label{Van_der_Pole}
\begin{array}{l}
\dot{x}_{1}=x_{2}, \\
\dot{x}_{2}=-x_{1}-x_{2}+x_{2} x_{1}^{2} .
\end{array}
\end{equation}

A total of 209 trajectories of the system in (\ref{Van_der_Pole}) are sampled , each over the time interval $[0,10]$ with a sampling time of 0.02 seconds. The trajectories are recorded starting form initial conditions that are uniformly sampled from the set $[-1,1] \times [-1,1]$. The trajectories are then lifted to different dimensions, with $N=2,3,...,12$ for system identification.

Using the lifted trajectories, (\ref{A-determination}) is used to determine the system matrix $\hat{A}$. The identified system matrix is used to simulate the identified system from $t_0=0$ to $t_f=20$ seconds. The trajectories generated by the identified system are truncated to the first two dimensions and compared with the recorded trajectories of the nonlinear system in (\ref{Van_der_Pole}). For comparison, trajectories of the linear system obtained using the model-based Carleman linearization in \cite{SCC.Amini.ea2021} are also generated. Note that the system matrix $\hat{A}$ is computed directly using recorded data from the nonlinear system. The model-based Carleman linearization from \cite{SCC.Amini.ea2021} is used for comparison and analysis purposes only.

The resulting trajectories of the nonlinear system, the model-based Carleman linearized system, and the identified linear systems are shown in Figures \ref{fig:order_2}-\ref{fig:order_11}. To quantify the performance, the errors $\sup _{t \in [0,\tau^*]}\left \|x(t)-\hat z\mid_2(t)\right \|$ between the measured trajectories and the identified system trajectories, and $\sup _{t \in [0,\tau^*]}\left \|x(t)- z\mid_2(t)\right \|$, between the measured trajectories and the model-based Carleman linearized trajectories, are plotted in Figure \ref{fig:error_plot}. The parameters introduced in Sections \ref{section:Problem Statement} and \ref{section:Error Analysis and Prescribed Error Approximation} are selected as $ M = 1.5$, $R=4.1$, $C=33.7$, $\overline{N} = 13$ and $C_0=0.001$, all of which satisfy the conditions in Section \ref{section:Error Analysis and Prescribed Error Approximation} and in (\ref{decay_rate}). Using (\ref{error_bound}) the selected parameters can be seen to guarantee the error bound $\left \|x(t)- \hat{z}\mid_2(t)\right \|\leq 3$ over the time interval $t\in[0,0.2]$.

\remark{Since the algorithm returns $\hat{A}$ using $N^*$ or returns "Failed", insights into the system identification method and the error bounds for $N\neq N^*$ are not obtainable. To that end, this paper presents results for identified systems using lifting orders other than $N^*$ and compares the guaranteed error bounds and the error for a range of lifting orders.}

Figure \ref{fig:error_bound} illustrates the data-driven bound on $\sup _{t \in [0,\tau^*]}\left \|x(t)- \hat{z}\mid_2(t)\right \|$, developed in (\ref{error_bound}), compared with the model-based bound on $\sup _{t \in [0,\tau^*]}\left \|x(t)- z\mid_2(t)\right \|$, developed in \cite{SCC.Amini.ea2021}.

\section{Discussion}
\label{section:Discussion}

The results in Figures \ref{fig:order_2}-\ref{fig:error_plot} show that for the model-based Carleman linearized system, increasing the order $N$ produces a more accurate approximation of (\ref{Van_der_Pole}). However, as seen in Figure \ref{fig:order_11} for the data-driven identified system, increasing the order $N$ past a certain point can result in an identified model that is inaccurate. As indicated by the bound in (\ref{error_bound}), as long as $\tau^*$ is small enough to satisfy (\ref{tau_bound}), the Carleman linearization is guaranteed to get better with increasing truncation order. However, when used in conjunction with a system identification method, there is a critical truncation order after which the estimation becomes less accurate.  

The optimal truncation order can be estimated using the bound developed in (\ref{error_bound}), as illustrated by Figures \ref{fig:error_plot} and \ref{fig:error_bound}. Figure \ref{fig:error_plot}  shows that in the numerical experiment, $N=6$ produces the identified system with the lowest error, while the analytical bound, (see Figure \ref{fig:error_bound}) indicates that $N^*=3$, meaning for $N=3$ the smallest guaranteed error bound is obtained. Since the analytical bound is conservative, some discrepancy between the numerical results and the analytical bound is expected.

To further explore the discrepancy, it is instructive to plot the error between the full trajectories of the model-based Carleman linearized system and the data-driven identified linear system (see Figure \ref{fig:error_plot_full_z_zhat}) and the error between the same trajectories, truncated to the first two dimensions (see Figure \ref{fig:error_plot_1_2_z_zhat}). The results indicate that while the full state estimation error $\sup _{0 \leq t \leq \tau^*}\left\|z(t)-\hat{z}(t)\right\|$ increases monotonically with $N$, the truncated state estimation error $\sup _{0 \leq t \leq \tau^*}\left\|z\mid_2(t)-\hat{z}\mid_2(t)\right\|$ initially decreases and then increases with increasing $N$. The analysis presented in this paper, that produces $N^*=3$, is based on a bound on the full state estimation error, which is observed to be an overly conservative bound on the truncated state estimation error for a subset of truncation orders. The authors postulate that development of an error bound for the truncated state estimation error will reduce the discrepancy between the analytical and the experimental results.


	
	\section{Conclusion and future work} 
	\label{section:Conclusion and future work}
	A system identification method using Carleman linearization is developed to identify a lifted linear system that produces trajectories which remain within a guaranteed error bound from the trajectory of a nonlinear system under mild assumptions. The
effectiveness of the technique is demonstrated by identifying an
approximation of the Van der Pol oscillator from data. The time interval $[0,\tau^*]$ over which the error bound is guaranteed is determined by the decay rate in the assumptions. It would be desirable for $\tau^*$ to be defined by the user, this would part of future work drawing from the results presented in \cite{scc.amini2022}.
 Analytical estimation of truncated state estimation error and the bound in Assumption \ref{as:4}, investigation of machine precision effects at high truncation orders, and development of a less conservative error bounds for the states of interest are part of future work.

\bibliographystyle{IEEETrans.bst}
\bibliography{ref}

\end{document}